\documentclass[10pt,journal,cspaper,compsoc]{IEEEtran}

\usepackage{amsfonts,amsmath, amssymb, graphicx, cite}

\newtheorem{client}{Client}
\newtheorem{challenge}{Challenge}
\newtheorem{solution}{Solution}
\newtheorem{val}{Value}

\begin{document}

\title{Cognitive Coordination of \\Global Service Delivery%
\thanks{
  This work is based in part on a paper presented at the 12th International Research Symposium on Service Excellence in 
  Management (QUIS12) \cite{VarshneyO2011}.}
}

\author{Lav~R.~Varshney, Shivali Agarwal, Yi-Min Chee, Renuka R.~Sindhgatta, \\Daniel V. Oppenheim, Juhnyoung Lee, and Krishna Ratakonda%
\thanks{L.~R.~Varshney was with the IBM Thomas J.\ Watson Research Center.  He is now with the Department of Electrical and Computer Engineering and the Coordinated Science Laboratory, University of Illinois at Urbana-Champaign, Urbana, IL (e-mail: varshney@illinois.edu).}
\thanks{S.~Agarwal and R.~Sindhgatta are with the IBM India Research Laboratory, Bangalore, India (e-mail: \{shivaaga, renuka.sr\}@in.ibm.com).}
\thanks{Y.-M.~Chee, D.~V.~Oppenheim, J.~Lee, and K.~Ratakonda are with the IBM Thomas J.\ Watson Research Center, Yorktown Heights, NY (e-mail: \{ymchee, music, jyl, ratakond\}@us.ibm.com).}
}

\maketitle
\thispagestyle{empty}
\pagestyle{empty}

\begin{abstract}
Formal coordination mechanisms are of growing importance as human-based service delivery becomes more globalized and informal 
mechanisms are no longer effective. Further it is becoming apparent that business environments, communication 
among distributed teams, and work performance are all subject to endogenous and exogenous uncertainty.

This paper describes a stochastic model of service requests in global service delivery and then puts forth a cognitive
approach for coordination in the face of uncertainty, based on a perception-action loop and receding horizon control.
Optimization algorithms used are a mix of myopic dynamic programming and constraint-based programming.
The coordination approach described has been deployed by a globally integrated enterprise in a very large-scale 
global delivery system and has been demonstrated to improve work efficiency by $10--15$\% as compared to manual 
planning.
\end{abstract}

\section{Introduction}
With the emergence of systems that bring together ubiquitous information technologies with the people and organizations 
they are transforming, it is important to understand how to direct and coordinate so as to achieve optimal efficiency.  
Firms are perhaps the most sophisticated of such sociotechnical systems, where people come together to develop 
innovative products and services.  The \emph{global service delivery} approach to doing knowledge work requires 
coordinating tens of thousands of specialized workers distributed around the world and has become prominent in 
many enterprises. Handling such large-scale agglomerations of people and machines, however, requires developing new 
abstractions, approaches, and algorithms.  This paper explicates the practical design of one such sociotechnical system 
for global service delivery and cognitive methods of coordination within it.  A key aspect of system design is
to understand and model the diversity of humans, and their preferences.

Whether engaged in designing a physical system like an airplane or building an information system like customer 
relationship management software, organizations providing informational services are becoming more and more 
globalized with an increasing degree of workforce specialization \cite{Bollier2011,MaloneLJ2011,Palmisano2006}:
specialized teams that concentrate on a narrow set of tasks can often be more productive than teams that are 
jacks-of-all-trades.  Indeed, the tradeoff between productivity benefits provided by specialization and coordination 
costs incurred with a distributed workforce are well-known in economic theory \cite{BeckerM1992, EhretW2010},
but globalization makes the value of specialization through division of labor more important 
now than ever before \cite{Stigler1951}. 

Unfortunately project failures, excessive delays, and significant financial losses have been observed in many global 
service delivery projects. Traditional project management techniques for co-located teams such as 
mutual adjustment through informal communication \cite{Mintzberg1989} do not scale well to a global workforce 
\cite{Gumm2006}. Four main problems in global software development include \cite{Wiredu2006}:
\begin{itemize}
  \item Conflicts of interest arising due to distributed work teams with local incentives,
  \item Interdependencies arising from distributed work processes,
  \item Technology representation problems arising from distributed technologies with local standards, and
  \item Uncertainties and equivocalities arising due to geographically and organizationally distributed information.
\end{itemize}
Although resolving conflicts of interest is certainly important, the cooperative elements of global collaboration 
are distinct from the coordinative ones \cite{GulatiWZ2012}.  As part of designing our cognitive coordination 
for global service delivery, we aim to minimize occurrence of the last three problems.

\begin{figure}
  \centering
  \includegraphics[width=3.5in]{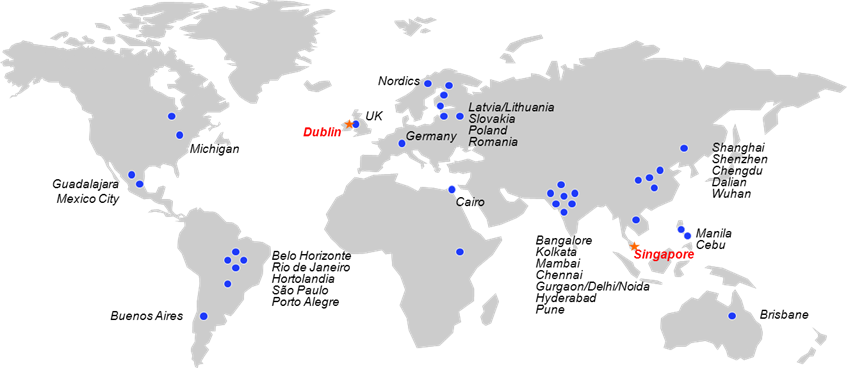}
  \caption{Delivery centers (dots) and coordination hubs (stars) distributed around the world.}
  \label{fig:map}
\end{figure}

The framework, approach, and algorithms detailed in the sequel arose from designing and implementing a new 
information technology framework for global service delivery: IBM's Application Assembly Automation (AAO), 
which has become a key component of IBM's Globally Integrated Capabilities \cite{IBM2009}. Large software 
development projects that were once carried out by large colocated teams are now 
broken into pieces and executed in isolation by an interchangeable delivery center.  Different delivery centers 
specialize in different aspects of software development, such as design, service-oriented architecture development, 
or testing, and are strategically located globally, as depicted in Figure~\ref{fig:map}. Each piece of work is 
routed to a delivery center through a construct called a \emph{work packet}, and the overall deliverables are 
then integrated in coordination hubs.  AAO has some similarities to other global delivery systems \cite{UptonF2005}.

\begin{figure}
  \centering
  \includegraphics[width=3.5in]{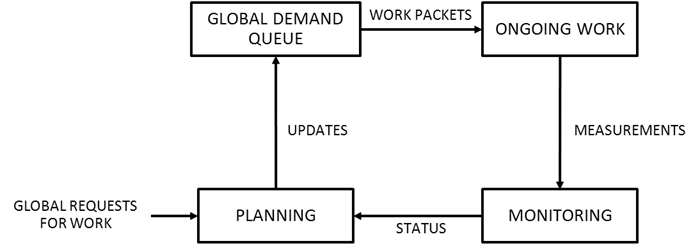}
  \caption{Perception-action cycle for global service delivery.}
  \label{fig:PA_cycle}
\end{figure}

The basic cognitive coordination approach we develop herein employs a \emph{perception-action loop} as a central 
construct, see Figure~\ref{fig:PA_cycle} for a block diagram representation.  
By perception-action loop, we mean the ability of a system to continuously monitor its own behavior
and the environment and to react accordingly to achieve a goal.  Such loops are not only useful for describing 
human cognition \cite{KlyubinPN2007}, but also for building cognitive dynamic systems \cite{Haykin2012,Haykin2014,HaykinF2014}.
Although cognitive coordination can certainly support human decision making \cite{OppenheimBRC2011a},
automated assignment of work to workers via task lists is considered here.  

The coordination algorithms follow the principle of receding horizon control (RHC).  In RHC, also known as model predictive control, 
an optimization problem is solved at each time step to determine a plan of action over a fixed time horizon.  The 
first control action from this plan is applied to the system. Then, at the next time step the planning process is 
repeated with a new optimization problem created with the time horizon shifted one time step forward.  Optimization 
takes uncertainties and estimates of future quantities based on available information into account at each time step \cite{MattingleyWB2011}.  
The specific algorithm developed for computational efficiency in optimizing the large-scale system is based on a
Markov decision process (MDP) formulation, which yields a mix of myopic dynamic programming (which uses limited 
information) and of constraint-based programming (which uses heuristic stopping rules).

Cognitive coordination for global service delivery enables scaling to larger and larger numbers
of workers carrying out more and more work, faster response to business needs, and greater visibility.  It has also 
led to $10$--$15$\% improved quality and productivity based on initial findings.

\section{Global Service Delivery Basics and Technologies}

The basic idea of global service delivery is to undertake several large service engagements with a globally
distributed workforce.  

Traditional approaches to distributed service delivery have used deterministic models of the business environment, 
of communication among people, and of the work itself to address interdependencies. This has led to standardized 
communication protocols and encapsulations of service work, as well as coordination mechanisms that deal with interdependencies 
using business process and business entity lifecycles \cite{DesaiCS2009, OppenheimBRC2011, LeymannR2000, NigamC2003}.
Rather than strict business process management approaches, we use an instantiation of the \emph{work-as-a-service} protocol and algebra 
\cite{OppenheimVC2011,OppenheimVC2014} to encapsulate work into pieces and define operations for its management; 
this flexible protocol is amenable to handling uncertainties that are inherent in human-intensive work.  

Work-as-a-service formalizes various operations such as \emph{merge}, \emph{tear}, 
\emph{pause}, and \emph{resume} which allow the control actions we will need to develop the RHC algorithms 
\cite{OppenheimVC2011,OppenheimVC2014,VaculinCOV2012}. We do not go into details of the work-as-a-service 
algebra in this paper, but focus on the larger optimized coordination enabled by it.
Large service engagements can be decomposed into several smaller work packets, 
and conversely several work packets can be combined into larger work packets. These work packets can also be 
delegated and reassigned to other service providers without any global impact since they are self-contained 
and explicitly include dependency relationships. Since work packets of any level of specificity can be decomposed, 
delegated, and reassigned, any service 
engagement can be thought of as comprising several atomic service requests. 
A complete service engagement would have a (perhaps hierarchical) network of atomic service requests, each 
dealt with in the same manner (due to uniformity of work packets).

As part of developing a perception-action loop for global service delivery, it is important to understand 
timescales over which management and planning actions can be taken.  In our view there are four such basic 
timescales with associated actions:
\begin{enumerate}
  \item \emph{scale of years}: High-level strategy such as which country to locate a work center in response to 
	labor markets, costs, etc., as well as high-level strategy on kinds of service work to pursue.
  \item \emph{scale of months}: Hiring new people and dropping current people in response to gaps/gluts, as well 
	as decisions to pursue specific service engagements.
  \item \emph{scale of days/hours}: Assignment of work tasks to workers in response to needs, skills, synergies, 
	and interdependencies.
  \item \emph{scale of minutes}: Ad hoc rejiggering of work assignments in response to perturbations that cannot 
	be dealt with through re-planning
\end{enumerate}
Our main focus in this paper is on the day/hour scale and thus on assignment of work tasks to workers.

Besides the ability to act, it is important to measure what is going on within the system and in the external 
environment \cite{Kuhn1961}.  Moving from measurement definition to measurement collection in a global 
service delivery environment has oft been complicated.
Moreover, there have been inconsistencies from project to project of what gets collected, how it gets collected, and when.
Inconsistencies even occur in a given metric when measured across two executions of the same process
because of individual human variation.  

To address these measurement challenges, we introduced a metrics framework 
(detailed elsewhere \cite{OppenheimCV2012}). The framework provides consistency and commonality across comparable measurements; the ability to define 
arbitrary levels of granularity of what is being measured; flexibility of changing metrics in the face of contextual
changes; deep visibility at all levels; and automation.  

\begin{figure}
  \centering
  \includegraphics[width=3.5in]{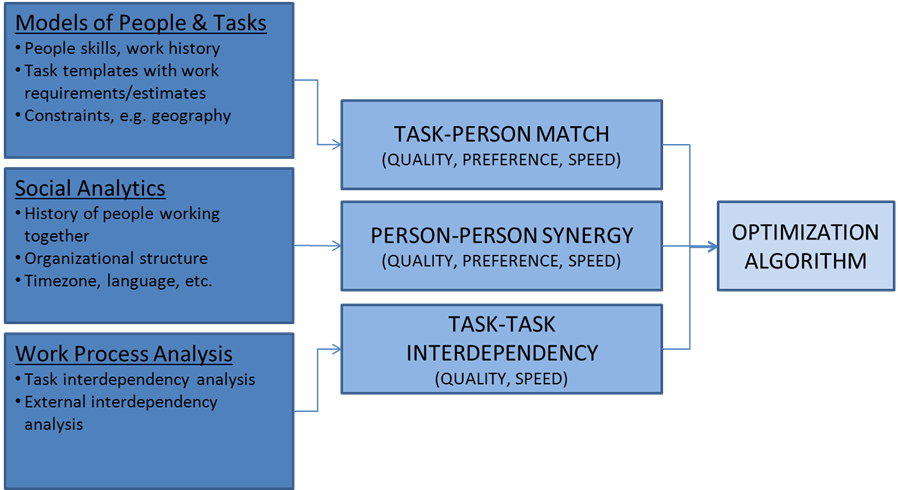}
  \caption{Planning combines data from multiple sources to generate constraints/objectives for optimal assignments.}
  \label{fig:data_sources}
\end{figure}

Besides its central role in continuous monitoring of system state, the metrics framework has also been used to 
characterize the work itself (encapsulated in terms of the work packet algebra), as well as to characterize the 
people that carry it out.  In particular, when developing measurements for global service delivery it is important 
to understand a skills assessment of people \cite{VarshneyWMFB2013}, an assessment of the work itself \cite{LiK2013}, 
as well as the social history of people \cite{LiuASL2013}.  Figure~\ref{fig:data_sources} illustrates how we use 
various measured data sources to inform optimization algorithms by understanding interdependencies in work, how well 
certain people are matched to certain tasks, and how well people work together.

\begin{figure*}
  \centering
  \includegraphics[width=7in]{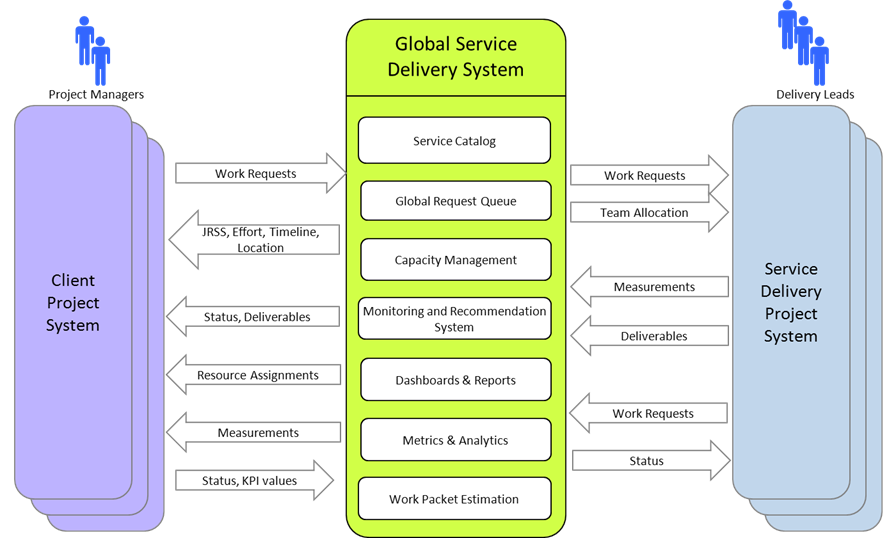}
  \caption{Global service delivery system design, with a coordination hub acting between requests and delivery centers.}
  \label{fig:system}
\end{figure*}

In addition to knowing what can be perceived and what can be acted upon, we also need to define a mechanism for
action.  Following fruitful precedent in controlling large-scale systems \cite{GrahamBK2009}, we employ middleware
technology to act as a coordination hub between service requesters and service providers, as depicted in Figure~\ref{fig:system}.
The goal of the middleware is to allow organizational modularity without incurring transaction costs 
\cite{BrusoniP2001,Foss2001,Baldwin2008}.  As can be noted, requesters of work need not interact with the
providers of work; all communication is routed through the coordination hub through formal mechanisms.  The
various information flows follow the work-as-a-service protocol.

\section{Perception-Action Loop}
\label{peractloop}
Our approach to coordinating sociotechnical systems is cognitive and uses the principle of receding horizon control, 
yet is similar to other coordination mechanisms \cite{MaloneC1994}.
A schematic diagram of the perception-action loop for global service delivery is depicted in Figure~\ref{fig:PA_cycle}.
The main exogenous perturbation is the introduction of new global requests for work.  These
are given to the planning subsystem, which also has access to a real-time status signal from the monitoring subsystem. 
Work is buffered in the global demand queue and is dispatched as work packets to workers.

\subsection{Receding Horizon Control}

RHC is a feedback control technique that became popular in the 1980s for physical systems \cite{MattingleyWB2011}, 
but does not seem to have previously been used for coordinating large-scale sociotechnical systems.  Thinking of time 
proceeding in hour-long steps, with RHC, an optimization problem is solved at each time step to determine a plan of 
work assignment over a fixed time horizon thereafter.  Optimization takes into account uncertainty and estimates of the future 
using available information at each time step. 

Since global knowledge work is characterized by a high degree of unpredictability from human factors, complexity, size, 
changing requirements, and the environment itself, a stochastic model is necessary. 
The nonlinear coordination policy uses feedback from real-time measurements and handles input
constraints, output constraints, and various control objectives.

Consider a work unit to be assigned as represented by a work breakdown structure such that each node in that 
structure graph represents a task to be assigned to one worker, and the edges represent interdependency constraints.  
There are many such units to be planned at a time. Work assignment is a matching problem where a work unit is assigned 
to an appropriately skilled resource so service delivery objectives are met. 
Existing work may get modified as time progresses and new work keeps arriving. The work units may undergo modifications 
of various types like altering the work structure or effort estimate or desired start/completion times or the preferred 
resources/geography. 

In addition to work dynamism, worker resources are also dynamic in terms of availability, skills, etc. 
A worker may go on leave and so work assigned to him may have to be reassigned. The worker may want to enhance his skills 
to eventually undertake different tasks than before. The worker's role may change, e.g.\ from developer to subject matter
expert, which will necessitate reassignment of tasks planned assuming his previous role.

With this inherent dynamism at various timescales of control, RHC is useful for work assignment in global delivery systems. 
The need becomes even more pronounced due to the volume of work and the scale of the teams. The output of 
planning is a \emph{work system plan} that is continuously updated in a palimpsestic manner
as output of the planning subsystem in Figure~\ref{fig:PA_cycle}.  The work plan decision variables are actions at each
timescale of action:
\begin{itemize}
  \item \emph{L1 decision}, e.g.\ start work center in Wisconsin next year;
  \item \emph{L2 decision}, e.g.\ hire 13 Java programmers next month;
  \item \emph{L3 decision}, e.g.\ assign task 347 to worker 872 to start tomorrow; and
  \item \emph{Brownout decision}, e.g.\ reassign task 85 from worker 872 to worker 873 since worker 872 just got ill.
\end{itemize}
For psychological benefits, these decisions are translated into task lists before presentation to workers \cite{BellottiDGFBD2004}.

The remainder of this section discusses algorithmic approaches for dynamic optimization,
focusing on L3 decisions: assigning work to workers.

\subsection{Markov Decision Process Formulation}
\begin{figure}
  \centering
  \includegraphics[width=3.5in]{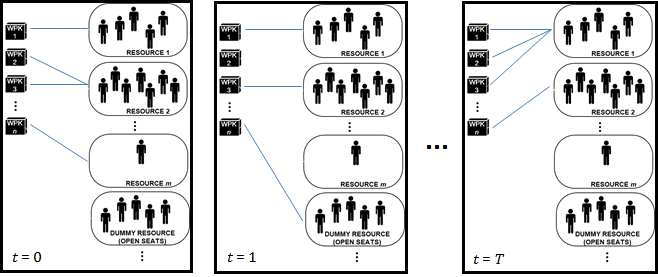}
  \caption{Depicting the Markov decision process view of work assignment for global service delivery over time $t$.  Note that 
	the possible dummy resources can be added through L2 actions, whereas assignment is through L3 actions (blue lines).}
  \label{fig:mdp}
\end{figure}

When focusing on L3 decisions (the assignment of work to workers) and certain aspects of 
L2 decisions on hiring/releasing workers, a Markov decision process (MDP) formulation is natural,
since time proceeds in stages and there is a notion of state that captures all dependencies
between past and future \cite{Howard1960}.  As depicted in Figure~\ref{fig:mdp}, at each stage
there is an assignment to be made, along with a longer-term decision on whether to invoke
``dummy resources'' that correspond to resources that have not yet been hired.  It is assumed
that preemption is not allowed due to the loss of robustness and the inefficiency \cite{OppenheimVC2014}
it causes.  Hence once a task is assigned to a worker, the state variable of that worker is set
to the time remaining for the task.

The goal of a coordination algorithm is to optimize assignments within the MDP problem, however without preemption this
is a computationally complicated integer program.  A natural approach for finding the globally optimal solution would
be via dynamic programming, but the state space is incredibly large.  In our global delivery setting, we foresaw 
requirements of: (a) small-scale optimization over short timescales for every invocation almost at the time granularity 
of every deliverable arrival, and (b) approximate global optimization periodically so as to keep things on track.
An RHC approach may therefore be effective.

Algorithmic development is built on the hypothesis that a myopic version of dynamic programming with a 
limited state space will be effective in cases of small and frequent perturbations whereas a constraint-based programming
approach that uses global information heuristically\footnote{Heuristic stopping rules to limit computational complexity
are included in software packages such as ILOG ({\tt http://www-01.ibm.com/software/info/ilog/}).} will be effective 
to find assignments given a fresh input set of large size.

We develop coordination that combines these two approaches: 
constraint-based programming uses global information but a heuristic optimization principle, whereas 
limited-horizon dynamic programming uses limited information but performs full optimization.
Thus we end up with RHC that has stage-by-stage bipartite matching within the
myopic dynamic programming and periodic globally optimal \emph{scheduling checkpoints} via constraints.
Though we omit formal statement and proof of this result, one can argue  that 
it is nearly optimal to use the (optimal) Hungarian method for bipartite matching and constrained programming 
for (nearly) globally optimal checkpoints, by closely examining Bellman's equation.  We present experiments to adjudge 
performance in Section~\ref{sec:simul}. 

\subsection{Constraint-Based Programming}
To get near-optimal scheduling checkpoints that use information far into the future, we use 
constraint-based programming \cite{NavehRAGC2007,AsafERCGOM2010}. Inputs are a set of work units and the pool of resources,
recast as a set of constraints and objectives.  The output of optimization is a complete schedule of the work 
as assigned on worker calendars. The main constraints from Figure~\ref{fig:data_sources} are as follows.

{\bf Skill match}: To do work, some skills and attributes are mandatory whereas others are optional.  For a government services 
programming task, Java skill and American geographical location may be mandatory whereas knowledge of tax codes may be optional.
For matching work with mandatory requirements, the algorithm discards resources that do not meet hard constraints
like skill, role, or location.  Once the eligible set of resources is obtained, an affinity score with respect to optional factors 
like project, application, tools used, or account is determined from the encapsulated information in the work packet and 
information maintained about each worker on expertise and experience.  There is an affinity score for
each resource and work packet pair, $\langle$res, wpk$\rangle$.

{\bf Time distribution of resources amongst tasks}: There may be work policies that dictate time allocation.  For example, a policy may 
require only one task to be performed at a time whereas other policies may allow resources to perform tasks in parallel.

{\bf Dependency among tasks within a work unit:} Typically, projects have dependencies expressed as partial ordering constraints like 
\emph{start-to-finish}, \emph{finish-to-start}, \emph{start-to-start}, and \emph{finish-to-finish}. 

{\bf Resource availability}: Constraints are needed to account for available time when planning for new work and may include the list of holidays 
for a resource.

The work to be assigned can be in-progress, starting-in-near-future, or far-in-future. The algorithm is aware of these time 
attributes and accordingly modifies plans. Work that is starting in the near future should undergo minimum adjustments in plan
since it is psychologically important for workers to have some idea of what work is coming next in their task list.
The temporal stability of the algorithm should, however, be parametrized to support cases where great dynamism is appropriate.

Robustness is also important for global service delivery so service requesters do not feel the impact of the perturbations happening within 
the delivery system; the system should absorb internal perturbations without hampering customer commitments and service level agreements. 
Indeed, if the algorithm frequently suggests many changes in the work assignment, then it may be difficult for delivery managers to 
make a commitment to customers and put forth a plan for each deliverable. 

An interesting aspect of the constraint-based formulation is in the objectives, which have an inherent tension among them. 
The objectives for the real-time, large-scale work assignment are:
\begin{enumerate}
  \item Work should be completed by the deadline. This should be based on the priority of the work.
  \item Resources with the best possible skill match should be chosen for each task. 
  \item Adjust the plan to accommodate dynamic changes to work or resources such that the properties of stability and robustness are maintained.
\end{enumerate}
(Note the inherent tension with the first objective, since the best resources may be engaged in other work.) 

\subsubsection{Constraint-based program}
Now let us mathematize the optimization problem.  The inputs are as follows:
\begin{itemize}
  \item $N$, the number of work packets, $i\in\{1,\ldots,N\}$
  \item $M$, the number of resources, $j\in\{1,\ldots,M\}$
  \item $R$, the number of deliverables, $k\in\{1,\ldots,R\}$
  \item $\sigma_{ij}$, an input matrix where an entry is a score if location and role of $i$ and $j$ match and zero otherwise
  \item $f_i$, the effort for packet $i$
  \item $C_k$, the committed end date for deliverable $k$: this may be empty for fresh ones
  \item $St_k$, the input start date for deliverable $k$
  \item $p\in\{1,\ldots,P\}$, an element in the set of priorities
  \item $\pi_p$, the penalty of scheduling later than start date: the penalty with high priority can be set very large, e.g.\ it can be exponential in priority to model the objective function
  \item $P_{kp}$, a matrix such that an entry is $1$ if deliverable $k$ has priority $p$ and 0 otherwise
  \item $\tau \in \{1,\ldots,T\}$, the time for assignment
  \item $A_{j\tau}$, a matrix derived from resource calendar
  \item $z_{ik}$, is $1$ if deliverable $k$ contains the packet $i$
  \item $d_{ii'}$, the dependency type between packets $i$ and $i'$
\end{itemize}
The decision variables are as follows:
\begin{itemize}
  \item $x_{ij}$, which is true if packet $i$ is assigned to resource $j$, and false otherwise
  \item $e_{i}$, the end date of packet $i$
  \item $s_{i}$, the start date of packet $i$
  \item $E_{k}$, the end date of deliverable $k$
  \item $S_{k}$, the start date of packet $k$
  \item $y_{j\tau}$, which is true if resource $j$ works at time $\tau$
\end{itemize}

This leads to the following optimization objectives.
\begin{align*}
&\min \sum_{k,p}\pi_p P_{kp} (E_k - St_k) \quad\mbox{{\sc{\tiny(Objective 1)}}} \\ \notag
&\max \sum_{i,j}\sigma_{ij} x_{ij} \quad\mbox{{\sc{\tiny(Objective 2)}}} \\ \notag
\end{align*}
where the two objectives can be combined into a single objective Lagrangian by subtraction.
The following constraints are also imposed.
\begin{align*}
&\mbox{s.t.} \\ \notag
&\quad \sum_{i,j} x_{ij} \sigma_{ij} > 0 \quad\mbox{{\sc{\tiny(matching location and role)}}}\\ \notag
&\quad \sum_{j} x_{ij}=1 \mbox{for all} i  \quad\mbox{{\sc{\tiny(one packet to one resource)}}}\\ \notag
&\quad \sum_{j} x_{ij}(e_i - s_i)\geq f_i \mbox{ for all } i  \quad\mbox{{\sc{\tiny(planned end date accounts for effort)}}}\\ \notag 
&\quad \sum_{j} (x_{ij} x_{i'j}=1)\Rightarrow (e_i<s_i'\vee s_i>e_i') \mbox{ for all } i\neq i'\\ \notag
&\hspace{2in}\mbox{{\sc{\tiny(one packet at a time)}}}\\ \notag
&\quad \sum_{ii'jj'} d_{ii'}(x_{ij} x_{i'j'}=1)\Rightarrow e_i<s_i'   \quad\mbox{{\sc{\tiny(sequential dependency)}}}\\ \notag
&\quad S_k\geq St_k \quad\mbox{{\sc{\tiny(planned start date after input start date)}}}\\ \notag
&\quad C_k\neq 0 \Rightarrow E_k\leq C_k \quad\mbox{{\sc{\tiny(end date obeys committed date)}}}\\ \notag
&\quad E_k-S_k=\sum_{i,k}z_{ik}(e_i-s_i)\quad\mbox{{\sc{\tiny(deliverable consists of work packets)}}}\\ \notag
&\quad \sum_{j,\tau} (y_{j\tau} x_{ij}A_{j\tau})=f_i \mbox{ for all } i \\ \notag
&\quad x_{ij}\in\{0,1\}
\end{align*}

This is a combinatorial optimization problem, so due to computational complexity constraints, one must 
use heuristic stopping criteria.

\subsection{Dynamic Programming}
With scheduling checkpoints established, a myopic form of dynamic programming that only looks a few time
steps ahead is used in between \cite{Bellman1957}.  Since the problem is one of assigning workers to
tasks, the affinity scores defined above (Figure~\ref{fig:data_sources}) are used as inputs to 
an implementation of the Hungarian method for bipartite matching \cite{Kuhn1955} for optimization 
within stages of the dynamic programming.  The Bellman principle is used for stage-by-stage optimization.
Note that there are strong similarities to queuing network control \cite{TehraniZ2010}.
The formalism is as follows.

The \emph{decision epoch} occurs at each work packet arrival, at each work packet completion 
indication, and at each information update on work completion estimate, with time instances indexed 
as $t=1,2,\ldots$. So time is event-driven.  The \emph{state space} has variables that specify 
how far along each worker is towards completing his/her current work packet.  The \emph{action space}
is the assignment matrix $A_t$ of work packets to workers.  For each pairing between work packets and workers, 
the \emph{reward} is determined by a cost $c_{ij}(t)$ that is computed from expertise match and the various 
other input data.  So the reward is $\sum_{i,j: i \rightarrow j} c_{ij}(t)$.  Since the goal is 
throughput/utilization, this cost will typically be the time required to complete work.  Due to lack of 
ability to preempt, $c_{ij}(t)$ is time-dependent and has action-dependence such that when things are assigned 
into the future, it introduces an infinite-valued entry for $c_{ij}(t)$.

The \emph{objective} is the long-term average of the cost functional:
\[
\lim_{T\rightarrow \infty} \frac{1}{T} \mathbb{E}\left[ \sum_{t=1}^T \sum_{i,j: i \rightarrow j} c_{ij}(t) \right]
\]
and the \emph{policy} is to perform optimal bipartite matching at every decision epoch (using the Hungarian method which runs 
in polynomial time): 
\[
A_t = \min_{i\rightarrow j} \sum_{i,j: i \rightarrow j} c_{ij}(t) \mbox{ for all } t = t_0, t_0 + 1, \ldots, T
\]
starting at the end horizon $T$ and working backwards using stochastic dynamic programming, while updating the action-dependent rewards.  
Results are placed into each workers task list.  

\section{Simulation Results}
\label{sec:simul}
\begin{figure*}
  \centering
  \includegraphics[width=5.5in]{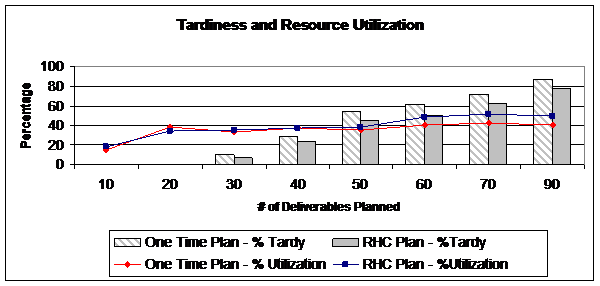}
  \caption{Comparing the performance of RHC coordination with typical human-oriented coordination.}
  \label{fig:results}
\end{figure*}
We designed an event-driven simulator to mimic global delivery work planning by having 
the work and resources instantiated in accordance with properties of real systems. The 
simulation engine generates deliverables with the desired details --– skills required, preferred 
date of completion, work packets and their dependences. The system is bootstrapped with a 
fixed number of resources whose skill profiles match with the incoming work. Upon arrival of 
new deliverable requests, assignment of work packets to resources is done based on the algorithms
described in Section~\ref{peractloop}.  After each assignment run of the optimization engine, 
resources have updated work lists and calendars.  

Simulation is carried out for two scenarios: 
SINGLE and RHC. In the SINGLE scenario, the plan once generated for deliverables does not 
undergo any changes. This is similar to manual project planning and scheduling processes, 
where planning occurs during the initial stages and does not change unless absolutely 
necessary. In the RHC scenario, there is re-planning 
at periodic intervals to ensure optimality by considering the latest set of deliverables. 
We define and measure two performance metrics to compare the two scenarios: percentage of deliverables 
meeting the deadline and percentage of resources utilized. 

Figure~\ref{fig:results} shows performance for both SINGLE and RHC. We create $25$ 
resources with varying skills and roles. We conduct eight simulation runs with varying number of deliverables. 
As shown, when the number of deliverables is low (10--30) or when there are 
abundant resources, the percentage of tardy deliverables is the same. The percentage utilization
of resources is also similar. However, as the number of deliverables increases, the percentage 
of tardy deliverables reduces and the resource utilization increases for the RHC scenario. 
Improved metrics indicate the improved efficiency of RHC as compared to SINGLE scenario. 
We observe up to 10\% improvement in utilization when RHC is adopted. 

An interesting threshold 
we observed was that when the number of deliverables is very large (90 or 1080 work packets), 
some of the resources reached utilization of $> 95\%$. This acted as a limit on further 
improvement due to RHC. However, with a different set of work definitions, it is possible 
to achieve up to 15\% improvement. 

Since the work definition used for simulation was taken from 
real projects, the results provide significant insights into the real system deployment,
discussed next.

\section{System Deployments}
In this section, we provide two brief case studies of some deployments of the system described in this work.

\subsection{Tire Manufacturer}
\begin{client}
World leader in manufacturing tires and related products. Project involved integration of disparate systems into cohesive 
Order-To-Cash (OTC) functionality across the client's enterprise. 
\end{client}
\begin{challenge}
Client needed to address specific business challenges to remain competitive.  In particular,
data across several legacy systems could not be shared efficiently, the technology environment could 
not support real-time or predictive information analysis, the supply chain could not be viewed 
and managed holistically, and there were high costs related to maintaining multiple legacy systems.
\end{challenge}
\begin{solution}
The service engagement therefore was to develop a service-oriented architecture based on an enterprise service bus,
so as to integrate various disparate systems and technologies (mainframe, DRP, SAP, Highjump, etc.) as well as 
legacy systems.  There was need to support all transformation logic at a centralized place in middleware 
and provide a standard method for extracting enterprise resource planning data into other systems for analysis. 

The solution was implemented using multiple IBM Global Delivery locations across the U.S., India, and China.
The multifarious, interdependent, and high-volume work was done by following the delivery methods and coordination 
mechanisms described herein.  Indeed, as part of delivery over $210$ technology interfaces were identified and 
developed. 
\end{solution}
\begin{val}
In leveraging global assets to improve productivity, the cognitive approach reduced coordination costs and allowed 
$10$\% reduction in effort hours.  For the client, this increased business efficiency through integration of business processes,
provided legacy synchronization of master and transactional data across divisions, and improved the client's
responsiveness to changing business needs.
\end{val}

\subsection{Telecommunications Company}

\begin{client}
Major telecommunications company which offers the local exchange carrier for telephone and 
DSL Internet services in most of Canada.  Project involved enabling a business model that improved performance and reduced cost in
performing data extraction, transformation and loading (ETL) via architecture standards and best practices for sustainable productivity improvements. 
\end{client}
\begin{challenge}
Client architecture met basic requirement of data movement but was unstructured and inefficient. The gaps in the architecture are 
also pervasive at the implementation level such as no standardized components, redundancy, and inflexible solutions.  Additionally, 
the client was unable to provide detailed implementation specifications due to their contractual agreements with their customers 
and vendors.
\end{challenge}
\begin{solution}
Based on initial analysis, IBM team developed client/environment specific recommendations and generic best practices based on the cognitive coordination 
approach for Data Stage and Tera Data. The engagement team created and presented communication plans catering to different groups 
both within the client organization and with vendors to enable ETL 2.0 recommendations. The business analytics and optimization
team performed analysis of various applications and downstream data integration methodologies to provide recommendations and 
solutions for efficiency. The team also recommended multiple approaches for scalability and maintenance by leveraging data virtualization 
in combination with data integration.
\end{solution}
\begin{val}
The cognitive coordination model helped in reducing the overall cost for the project through a shared delivery model, usage of assets and accelerators 
and collaboration with technical subject matter experts. The estimation model provided a $10$\% reduction in the overall 
effort level thereby helping the client meet tight timelines and budget. The reduction was achieved through leveraging reusable 
components, and parallel processing. Also, using the encapsulated templates, checklists and best practices, the project was able 
to achieve a significant reduction (more than $50$\%) in the number of defects. It successfully overcame the challenges of working 
across different time zones and multiple languages and overlapping waves to ensure smooth delivery. Finally, the concept of 
\emph{golden data client} was used to build production quality data ahead of time and get business user commitment and ownership for data.
\end{val}

\section{Conclusion}

Sociotechnical systems for delivering services are subject to various forms of uncertainty. 
Indeed, ``uncertainty is 
what typifies projects. It's the nature of the beast'' \cite{Goldratt1997}.  Though always present,
this inherent uncertainty is becoming more noticeable as inefficiencies are being squeezed out 
of service organizations. As has been noted, ``after years of optimizing supply chains, 
outsourcing, automation, and stripping costs and inefficiencies out of the back office, 
most employees spend very little of their day working on regularized activities. What they 
do is they manage exceptions to processes'' \cite{TapscottW2006}.

These issues are magnified in the \emph{global} service delivery context, where manual coordination
procedures have become inefficient in dealing with uncertainties in a scalable manner.
Formal coordination mechanisms that measure system state and take actions to respond are 
becoming crucial.

In this work, we have reported on our experience in coordinating a large-scale sociotechnical system
for global service delivery.  Using a cognitive coordination framework, a Markov decision process 
formulation, and computationally-implementable receding horizon control algorithms, we 
have developed a middleware deployment that achieves $10$--$15$\% improvement over existing
coordination approaches.  These improvements are measured not only in realistic simulation
studies, but also in client project deployments.  The basic frameworks, formulations, algorithms,
and technologies can serve as the basis for other similar problems of coordination.

\section*{Acknowledgment}
The authors thank Rong Liu and Bikram Sengupta for discussions, and colleagues in IBM Global Business Services for 
deploying the system described herein.

\bibliographystyle{IEEEtran} 
\bibliography{abrv,conf_abrv,global_delivery}

\begin{thebibliography}{10}
\providecommand{\url}[1]{#1}
\csname url@samestyle\endcsname
\providecommand{\newblock}{\relax}
\providecommand{\bibinfo}[2]{#2}
\providecommand{\BIBentrySTDinterwordspacing}{\spaceskip=0pt\relax}
\providecommand{\BIBentryALTinterwordstretchfactor}{4}
\providecommand{\BIBentryALTinterwordspacing}{\spaceskip=\fontdimen2\font plus
\BIBentryALTinterwordstretchfactor\fontdimen3\font minus
  \fontdimen4\font\relax}
\providecommand{\BIBforeignlanguage}[2]{{%
\expandafter\ifx\csname l@#1\endcsname\relax
\typeout{** WARNING: IEEEtran.bst: No hyphenation pattern has been}%
\typeout{** loaded for the language `#1'. Using the pattern for}%
\typeout{** the default language instead.}%
\else
\language=\csname l@#1\endcsname
\fi
#2}}
\providecommand{\BIBdecl}{\relax}
\BIBdecl

\bibitem{VarshneyO2011}
L.~R. Varshney and D.~V. Oppenheim, ``Coordinating global service delivery in
  the presence of uncertainty,'' in \emph{Proc. 12th Int. Research Symp.
  Service Excellence Manage. (QUIS12)}, Jun. 2011, pp. 1004--1014.

\bibitem{Bollier2011}
D.~Bollier, \emph{The Future of Work: What It Means for Individuals,
  Businesses, Markets and Governments}.\hskip 1em plus 0.5em minus 0.4em\relax
  Washington, DC: The Aspen Institute, 2011.

\bibitem{MaloneLJ2011}
T.~W. Malone, R.~J. Laubacher, and T.~Johns, ``The age of
  hyperspecialization,'' \emph{Harvard Bus. Rev.}, vol.~89, no. 7/8, pp.
  56--65, July-Aug. 2011.

\bibitem{Palmisano2006}
S.~J. Palmisano, ``The globally integrated enterprise,'' \emph{Foreign Aff.},
  vol.~85, no.~3, pp. 127--136, May-June 2006.

\bibitem{BeckerM1992}
G.~S. Becker and K.~M. Murphy, ``The division of labor, coordination costs, and
  knowledge,'' \emph{Quart. J. Econ.}, vol. 107, no.~4, pp. 1137--1160, Nov.
  1992.

\bibitem{EhretW2010}
M.~Ehret and J.~Wirtz, ``Division of labor between firms: Business services,
  non-ownership-value and the rise of the service economy,'' \emph{Service
  Sci.}, vol.~2, no.~3, pp. 136--145, Fall 2010.

\bibitem{Stigler1951}
G.~J. Stigler, ``The division of labor is limited by the extent of the
  market,'' \emph{J. Polit. Econ.}, vol.~59, no.~3, pp. 185--193, Jun. 1951.

\bibitem{Mintzberg1989}
H.~Mintzberg, \emph{Mintzberg on Management}.\hskip 1em plus 0.5em minus
  0.4em\relax New York: Free Press, 1989.

\bibitem{Gumm2006}
D.~C. Gumm, ``Distribution dimensions in software development projects: A
  taxonomy,'' \emph{{IEEE} Softw.}, vol.~23, no.~5, pp. 45--51, Sept.-Oct.
  2006.

\bibitem{Wiredu2006}
G.~O. Wiredu, ``A framework for the analysis of coordination in global software
  development,'' in \emph{Proc. 2006 Int. Workshop Global Softw. Dev.
  Practitioner}, May 2006, pp. 38--44.

\bibitem{GulatiWZ2012}
R.~Gulati, F.~Wohlgezogen, and P.~Zhelyazkov, ``The two facets of
  collaboration: Cooperation and coordination in strategic alliances,''
  \emph{Acad. Manage. Ann.}, vol.~6, no.~1, pp. 531--583, Jun. 2012.

\bibitem{IBM2009}
{IBM Global Business Services}, ``Application assembly optimization: A new
  approach to global delivery,'' Aug. 2009.

\bibitem{UptonF2005}
D.~M. Upton and V.~A. Fuller, ``Wipro technologies: The factory model,''
  Harvard Business School: 9-606-021, Oct. 2005.

\bibitem{KlyubinPN2007}
A.~S. Klyubin, D.~Polani, and C.~L. Nehaniv, ``Representations of space and
  time in the maximization of information flow in the perception-action loop,''
  \emph{Neural Comput.}, vol.~19, no.~9, pp. 2387--2432, Sep. 2007.

\bibitem{Haykin2012}
S.~Haykin, ``Cognitive dynamic systems: Radar, control, and radio,''
  \emph{Proc. {IEEE}}, vol. 100, no.~7, pp. 2095--2103, Jul. 2012.

\bibitem{Haykin2014}
------, ``Cognitive dynamic systems,'' \emph{Proc. {IEEE}}, vol. 102, no.~4,
  pp. 414--416, Apr. 2014.

\bibitem{HaykinF2014}
S.~Haykin and J.~M. Fuster, ``On cognitive dynamic systems: Cognitive
  neuroscience and engineering learning from each other,'' \emph{Proc. {IEEE}},
  vol. 102, no.~4, pp. 608--628, Apr. 2014.

\bibitem{OppenheimBRC2011a}
D.~Oppenheim, S.~Bagheri, K.~Ratakonda, and Y.-M. Chee, ``Coordinating
  distributed operations,'' in \emph{Service-Oriented Computing}, ser. Lecture
  Notes in Computer Science, E.~M. Maximilien, G.~Rossi, S.-T. Yuan, H.~Ludwig,
  and M.~Fantinato, Eds.\hskip 1em plus 0.5em minus 0.4em\relax Berlin:
  Springer, 2011, vol. 6568, pp. 213--224.

\bibitem{MattingleyWB2011}
J.~Mattingley, Y.~Wang, and S.~Boyd, ``Receding horizon control,'' \emph{{IEEE}
  Control Syst. Mag.}, vol.~31, no.~3, pp. 52--65, Jun. 2011.

\bibitem{DesaiCS2009}
N.~Desai, A.~K. Chopra, and M.~P. Singh, ``Amoeba: A methodology for modeling
  and evolving cross-organizational business processes,'' \emph{ACM Trans.
  Softw. Eng. Methodol.}, vol.~19, no.~2, p.~6, Oct. 2009.

\bibitem{OppenheimBRC2011}
D.~V. Oppenheim, S.~Bagheri, K.~Ratakonda, and Y.-M. Chee, ``Agility of
  enterprise operations across distributed organizations: A model of cross
  enterprise collaboration,'' in \emph{Proc. SRII Global Conf. 2011}, Mar.
  2011.

\bibitem{LeymannR2000}
F.~Leymann and D.~Roller, \emph{Production Workflow: Concepts and
  Techniques}.\hskip 1em plus 0.5em minus 0.4em\relax Upper Saddle River, NJ:
  Prentice Hall, 2000.

\bibitem{NigamC2003}
A.~Nigam and N.~S. Caswell, ``Business artifacts: An approach to operational
  specification,'' \emph{IBM Syst. J.}, vol.~42, no.~3, pp. 428--445, 2003.

\bibitem{OppenheimVC2011}
D.~V. Oppenheim, L.~R. Varshney, and Y.-M. Chee, ``Work as a service,'' in
  \emph{Service-Oriented Computing}, ser. Lecture Notes in Computer Science,
  G.~Kappel, Z.~Maamar, and H.~R. Motahari-Nezhad, Eds.\hskip 1em plus 0.5em
  minus 0.4em\relax Berlin: Springer, 2011, vol. 7084, pp. 669--678.

\bibitem{OppenheimVC2014}
------, ``Work as a service,'' in \emph{Advanced Web Services}, A.~Bouguettaya,
  Q.~Z. Sheng, , and F.~Daniel, Eds.\hskip 1em plus 0.5em minus 0.4em\relax
  Springer, 2014, pp. 409--430.

\bibitem{VaculinCOV2012}
R.~Vaculin, Y.-M. Chee, D.~V. Oppenheim, and L.~R. Varshney, ``Work as a
  service meta-model and protocol for adjustable visibility, coordination, and
  control,'' in \emph{Proc. SRII Global Conf. 2012}, Jul. 2012, pp. 90--99.

\bibitem{Kuhn1961}
T.~S. Kuhn, ``The function of measurement in modern physical science,''
  \emph{Isis}, vol.~52, no.~2, pp. 161--193, Jun. 1961.

\bibitem{OppenheimCV2012}
D.~V. Oppenheim, Y.-M. Chee, and L.~R. Varshney, ``Allegro: A metrics framework
  for globally distributed service delivery,'' in \emph{Proc. SRII Global Conf.
  2012}, Jul. 2012, pp. 461--469.

\bibitem{VarshneyWMFB2013}
J.~Wang, K.~R. Varshney, A.~Mojsilovi\'{c}, D.~Fang, and J.~H. Bauer,
  ``Expertise assessment with multi-cue semantic information,'' in \emph{Proc.
  2013 IEEE Int. Conf. Serv. Oper. Logist. Inform. (SOLI)}, Jul. 2013, pp.
  534--539.

\bibitem{LiK2013}
Y.~Li and K.~Katircioglu, ``Measuring and applying service request effort data
  in application management services,'' in \emph{Proc. 2013 IEEE Int. Conf.
  Services Comput. (SCC)}, Jun. 2013, pp. 352--359.

\bibitem{LiuASL2013}
R.~Liu, S.~Agarwal, R.~R. Sindhgatta, and J.~Lee, ``Accelerating collaboration
  in task assignment using a socially enhanced resource model,'' in
  \emph{Business Process Management}, ser. Lecture Notes in Computer Science,
  F.~Daniel, J.~Wang, and B.~Weber, Eds.\hskip 1em plus 0.5em minus 0.4em\relax
  Berlin: Springer, 2013, vol. 8094, pp. 251--258.

\bibitem{GrahamBK2009}
S.~Graham, G.~Baliga, and P.~R. Kumar, ``Abstractions, architecture,
  mechanisms, and a middleware for networked control,'' \emph{{IEEE} Trans.
  Autom. Control}, vol.~54, no.~7, pp. 1490--1503, Jul. 2009.

\bibitem{BrusoniP2001}
S.~Brusoni and A.~Prencipe, ``Unpacking the black box of modularity:
  Technologies, products and organizations,'' \emph{Ind. Corporate Change},
  vol.~10, no.~1, pp. 179--205, Mar. 2001.

\bibitem{Foss2001}
K.~Foss, ``Organizing technological interdependencies: a coordination
  perspective on the firm,'' \emph{Ind. Corporate Change}, vol.~10, no.~1, pp.
  151--178, Mar. 2001.

\bibitem{Baldwin2008}
C.~Y. Baldwin, ``Where do transactions come from? modularity, transactions, and
  the boundaries of firms,'' \emph{Ind. Corporate Change}, vol.~17, no.~1, pp.
  155--195, Feb. 2008.

\bibitem{MaloneC1994}
T.~W. Malone and K.~Crowston, ``The interdisciplinary study of coordination,''
  \emph{ACM Comput. Surv.}, vol.~26, no.~1, pp. 87--119, Mar. 1994.

\bibitem{BellottiDGFBD2004}
V.~Bellotti, B.~Dalal, N.~Good, P.~Flynn, D.~G. Bobrow, and N.~Ducheneaut,
  ``What a to-do: studies of task management towards the design of a personal
  task list manager,'' in \emph{Proc. SIGCHI Conf. Hum. Factors Comput. Syst.
  (CHI 2004)}, Apr. 2004, pp. 735--742.

\bibitem{Howard1960}
R.~A. Howard, \emph{Dynamic Programming and {M}arkov Processes}.\hskip 1em plus
  0.5em minus 0.4em\relax Cambridge, MA: MIT Press, 1960.

\bibitem{NavehRAGC2007}
Y.~Naveh, Y.~Richter, Y.~Altshuler, D.~L. Gresh, and D.~P. Connors, ``Workforce
  optimization: Identification and assignment of professional workers using
  constraint programming,'' \emph{IBM J. Res. Develop.}, vol.~51, no. 3/4, pp.
  263--279, May 2007.

\bibitem{AsafERCGOM2010}
S.~Asaf, H.~Eran, Y.~Richter, D.~P. Connors, D.~L. Gresh, J.~Ortega, and M.~J.
  Mcinnis, ``Applying constraint programming to identification and assignment
  of service professionals,'' in \emph{Principles and Practice of Constraint
  Programming}, ser. Lecture Notes in Computer Science, D.~Cohen, Ed.\hskip 1em
  plus 0.5em minus 0.4em\relax Berlin: Springer, 2010, vol. 6308, pp. 24--37.

\bibitem{TehraniZ2010}
P.~Tehrani and Q.~Zhao, ``Multichannel scheduling and its connection to
  queueing network control problem,'' in \emph{Proc. Mil. Commun. Conf. (MILCOM
  2010)}, Nov. 2010, pp. 482--486.

\bibitem{Kuhn1955}
H.~W. Kuhn, ``The {H}ungarian method for the assignment problem,'' \emph{Nav.
  Res. Logist. Q.}, vol.~2, no. 1-2, pp. 83--97, Mar. 1955.

\bibitem{Bellman1957}
R.~Bellman, \emph{Dynamic Programming}.\hskip 1em plus 0.5em minus 0.4em\relax
  Princeton: Princeton University Press, 1957.

\bibitem{Goldratt1997}
E.~M. Goldratt, \emph{Critical Chain}.\hskip 1em plus 0.5em minus 0.4em\relax
  North River Press, 1997.

\bibitem{TapscottW2006}
D.~Tapscott and A.~D. Williams, \emph{Wikinomics: How Mass Collaboration
  Changes Everything}, expanded~ed.\hskip 1em plus 0.5em minus 0.4em\relax New
  York: Portfolio Penguin, 2006.

\end{thebibliography}

\end{document}